\documentclass{article}

\PassOptionsToPackage{sort}{natbib}

\usepackage{microtype}
\usepackage{graphicx}
\usepackage{subcaption}
\usepackage{booktabs} 

\usepackage{hyperref}

\usepackage[preprint]{icml2026}

\usepackage{amsmath}
\usepackage{amssymb}
\usepackage{mathtools}
\usepackage{amsthm}
\usepackage{mymacros}
\usepackage{xurl}

\usepackage[capitalize,noabbrev]{cleveref}

\theoremstyle{plain}

\theoremstyle{definition}

\theoremstyle{remark}

\usepackage[textsize=tiny]{todonotes}

\icmltitlerunning{Threat Modeling Intimate Partner Abuse in Agentic Systems}

\begin{document}

\twocolumn[
  \icmltitle{Threat Modeling Intimate Partner Abuse in Agentic Systems}



  \icmlsetsymbol{equal}{*}

  \begin{icmlauthorlist}
    \icmlauthor{Natalie Grace Brigham}{uw}
    \icmlauthor{Eugene Bagdasarian}{umass}
    \icmlauthor{Franziska Roesner}{uw}
    \icmlauthor{Tadayoshi Kohno}{gu}
  \end{icmlauthorlist}

  \icmlaffiliation{uw}{University of Washington}
  \icmlaffiliation{umass}{University of Massachusetts Amherst}
  \icmlaffiliation{gu}{Georgetown University}

  \icmlcorrespondingauthor{Natalie Grace Brigham}{nbrigham@cs.washington.edu}


  \vskip 0.3in
]



\printAffiliationsAndNotice{}  

\begin{abstract}
  Agentic systems are AI-powered technologies capable of autonomous action within digital environments, such as sending emails or making reservations.
  They are deeply connected to external tools as well as users' data and resources, making them vulnerable to attacks which exploit their access and capabilities.
  In this work, we apply the technology-facilitated abuse threat model, in the context of intimate partner violence and surveillance, to these systems, revealing threats that require deliberate investigation and mitigation.
  We categorize attacks based on system ownership: (1) Abuser-as-owner attacks demonstrate how agentic systems make existing forms of technology-facilitated abuse far easier and more scalable.
  (2) Victim-survivor-as-owner attacks show how agentic systems create a new attack surface which abusers can exploit.
  In light of these risks, we outline research and development directions for effective mitigation strategies.

  \textit{Content warning: This paper includes content about intimate partner violence and surveillance.}
\end{abstract}

\section{Introduction}
From spyware~\cite{chatterjee2018spyware, almansoori2022global, tseng2020tools} to financial technologies~\cite{bellini2023digital}, IoT/smart-home devices~\cite{stephenson2023abuse}, and generative AI~\cite{bailey2021ai, kim2026ai}, abusers\footnote{While we acknowledge the reductive nature of this label, we use the term ``abuser'' for clarity, given that we address strategies of abusive technology use and following \citet{kim2026ai}} can broadly exploit or weaponize technology in intimate partner violence (IPV) and surveillance (IPS), phenomena affecting millions~\cite{zhang2026national, who2025violence}.
In light of this, the computer security and privacy research community has applied threat modeling techniques---systematic methods for identifying attacks and defenses---to technology-facilitated abuse (TFA) in intimate partner contexts.
In this work, we extend this practice to the emerging, rapidly developing class of agentic systems.

These systems, also called AI agents, are defined as ``compound software systems, inclusive of one or more AI models, that operate within an environment and take actions within it''~\cite{feng2025levels}. Although they are built on large language models (LLMs) to drive their functionality (e.g., to determine necessary actions to complete a task, generate necessary content, interact with the user), they extend purely conversational AI systems (i.e., early AI chatbots) given their ability to take actions within their environment. For example, a conversational AI system might be able to generate an email requested by the user but an agentic system has the ability, given a connection to the user’s email account, to send the email on behalf of a user.

Agentic systems may be highly connected to the world through external tools and interfaces as well as users' data and resources, enabling them to execute complex, sophisticated, and even technical tasks.
Given this access and capability, they are vulnerable to being used for malicious tasks such as cyberattacks~\cite{openai2026huggingface, andriushchenko2025agentharm, fang2024llmagentsautonomouslyexploit, fang2024llmagentsautonomouslyhack} or being manipulated, for example through prompt injection~\cite{debenedetti2024agentdojodynamicenvironmentevaluate, evtimov2025waspbenchmarkingwebagent, zhan2025injecagent}.

Increasingly, consumer AI systems that were initially purely conversational have integrated some agentic capabilities, especially internet search. For instance, ChatGPT introduced an ``agent mode'' and ``connectors'' to services like Gmail and Expedia that can expand the system's environment~\cite{openai_chatgpt_agent_mode}. Additionally, we are now seeing systems designed specifically to offer agentic capabilities, including with large amounts of autonomy and little user involvement or supervision.
OpenClaw, an open source agent, gained massive popularity in early 2026 with a reported 3.2 million monthly active users as of April 2026~\cite{openclawpopularity}.
In September 2026, Meta announced Muse, a personal assistant AI agent~\cite{metamuse}, and, later that month, OpenAI launched their agents, ``dots''~\cite{openAIdots}.
Through this, these major companies join a field of startups offering similar agentic products such as AGI-0~\cite{agiinc} and Nori AI~\cite{nori}.

Between new agentic systems being introduced and the trending of popular conversational systems towards agentic approaches, the importance of security and privacy considerations in such systems is evident.
Current research in this space predominantly assumes adversaries as \textbf{external actors}---such as a malicious website operator, advertiser, or other third party---or as an \textbf{untrustworthy component} of the system, such as compromised tools or data. 
However, \textbf{interpersonal adversaries}, including intimate partners, have not been equally considered, leading to oversight.
For example, security mechanisms such as risk assessments for tool calls may block the sharing of sensitive financial, medical or otherwise personal data with unknown third parties but not prevent it between, for instance, spouses, even though abusers can exploit such information.

In this work, we apply the technology-facilitated abuse threat model to identify risks related to IPV and IPS that require deliberate investigation and mitigation.
First, we present background on the threat model and its prior application to other technologies (Section~\ref{sec:background}) as well as our analysis methodology (Section~\ref{sec:methodology}). We then apply the model to agentic systems, first considering practical risks based on current or near-future capabilities in scenarios where the agent owner is the abuser (Section~\ref{sec:abuser-as-owner}) and where the victim-survivor is the owner (Section~\ref{sec:victim-survivor-as-owner}). Then, we raise potential or far-future risks (Section~\ref{sec:long-horizon}). Finally, we discuss recommendations (Section~\ref{sec:discussion}) and address alternative perspectives and counterarguments (Section~\ref{sec:alternative-views}).
\section{Background \& Related Work}
\label{sec:background}

IPV is a global public health and human rights crisis affecting millions worldwide, with estimates suggesting that one in four ever-partnered women aged 15-49 has experienced physical and/or sexual IPV during their lifetime~\cite{zhang2026national, who2025violence}. Within this landscape, technology-facilitated abuse (TFA), in which perpetrators use technology to surveil, coerce, or otherwise abuse victim-survivors, has emerged as a pervasive form of abuse~\cite{douglas2019technologyfacilitated, woodlock2017abuse}.
Only in the past decade has computer science research, mainly in human-computer interaction and security and privacy, begun to address these threats. 
Such research has revealed how traditional safety mechanisms, designed primarily to protect against external malicious actors, fail victim-survivors because they do not account for interpersonal threats, for example where the abuser leverages intimate knowledge of shared accounts and system mechanics to bypass standard controls~\cite{freed2018stalkers}.

\subsection{Technology-Facilitated Abuse Threat Model}
\label{sec:tfa-threat-model}

In this work, we consider technology-related threats posed to a victim-survivor of IPV or IPS by their abuser.
While our focus is on technology-facilitated abuse, we recognize that IPV encompasses various forms (e.g., physical, financial) which often intersect and interact. Consequently, interventions targeting digital abuse may introduce unintended tensions, such as potentially aggravating the abuser and escalating physical violence~\cite{freed2017digital}.

Under this threat model, we assume the abuser has or had intimate access to the target's data, devices, accounts, social networks, and/or physical space. This access is typically gained through ownership (e.g., holding a shared account or owning a device) or through compromise (e.g., coercing the victim-survivor for authentication credentials, guessing them, or covertly inspecting a device). Many attacks leverage this proximity while remaining relatively technologically mundane, with aims ranging from controlling a victim-survivor's access to digital and physical resources to surveilling and harassing them~\cite{freed2018stalkers}. Much prior work on IPV, IPS and technology~\cite{chatterjee2018spyware, tseng2020tools, zhang2025abusability, bellini2023digital, ceccio2023sneaky, daffalla2023asi} aligns with \citet{freed2018stalkers}'s model of a \textit{UI-bound adversary}---defined as ``an authenticated but adversarial user of a victim's device or account who carries out attacks by interacting with the standard user interface.''

Proximity is greatest when the abuser has physical control and diminishes as the victim-survivor escapes. However, factors such as financial dependence, shared children, or other binding circumstances may prevent a complete loss of contact. Furthermore, the stages of physical control, escape, and life apart may be experienced cyclically, with certain risks being acute during escape~\cite{matthews2017stories}.
Importantly, some attacks---including certain forms of surveillance and harassment---do not require direct access to the victim's device or accounts. Instead, remote channels such as phone calls or social media can be utilized~\cite{thomas2021sok, matthews2017stories}.

In some ways, agents may augment this threat model. For example, an abuser's use of an agent may allow them to execute more technically advanced attacks that were previously impossible for them (Section~\ref{sec:abuser-as-owner}) or, conversely, their access to a victim-survivor's agent may allow abusers to perform actions beyond what is achievable through conventional UI-based interaction alone (Section~\ref{sec:victim-survivor-as-owner}).

\subsection{Technology \& IPV}
\label{sec:tech-and-ipv}
\headingg{Applications.}
Prior work has investigated the ecosystem of consumer spyware mobile apps, including those of dual-use nature, that can be used in IPS~\cite{chatterjee2018spyware,almansoori2022global,tseng2020tools} and found these apps to introduce additional risks through their own data insecurity~\cite{liu2023no}.
Automation apps (e.g., OS Shortcuts, Samsung Modes \& Routines, Tasker, and IFTTT) can be used by abusers to monitor, impersonate, overload, and coerce victim-survivors~\cite{zhang2025abusability, cobb2020how}.
\citet{bellini2023digital} examine mobile banking and peer-to-peer payment apps, identifying threats related to compromised access due to physical proximity
as well as through remote means.

\heading{Devices.}
Commercial spy devices are easily used and frequently advertised for IPS~\cite{ceccio2023sneaky, tseng2020tools}. Everyday internet-connected home devices can be used for abuse through covert spying, unauthorized access, repurposing, and even intended use~\cite{stephenson2023abuse}.
Abusers often leverage physical and relational proximity to compromise victim-survivors' devices~\cite{tseng2020tools, bellini2023digital, matthews2017stories}.

\heading{Accounts.}
\citet{daffalla2023asi} and \citet{bhattacharya2026inconsistent} highlight how account security interfaces, often used to support victim-survivors in controlling their accounts, can be vulnerable to access spoofing and hiding attacks from abusers.
This vulnerability may, however, be mitigated through data exports~\cite{nonnenkamp2025hidden}.
Shared or linked accounts, such as those for cloud tools (e.g., iCloud) or phone plans, can also be exploited by abusers~\cite{tseng2020tools}.
Additionally, abusers may compromise credentials through coercion or tactics like guessing the answers to security questions~\cite{freed2017digital}.

\heading{AI.}
While to our knowledge no prior work considers the application of agentic systems for TFA, other works have considered conversational AI systems, such as chatbots.
\citet{kim2026ai} demonstrate how LLM chatbots can be used for coercive control across four scenarios. With the abuser as user, they show how the chatbots can generate content that (1) facilitates bullying and harassment, (2) can be used to coerce the victim-survivor, and (3) describes stalking tools and tactics learned during model training. With the victim-survivor as user, they consider a fourth scenario in which the abuser manipulates the system prompt to gaslight the user during their conversations with the chatbot.
\citet{suh2026psychologicalinfluencesconversationalai} posit that interaction with AI chatbots may cause psychological or physical harm when they fail to recognize and respond properly to a user in an abusive interpersonal relationship or perpetrating abuse.
\citet{ghoshal2026position} argue that AI companions pose IPV risks by facilitating the rehearsal of violent, coercive, or abusive behaviors without meaningful resistance, although the causal link between such interactions and actual abuse remains speculative.
Generative AI can also facilitate abuse through the creation of non-consensual intimate imagery~\cite{bailey2021ai}.
Though focused on external adversaries, \citet{peisze2026agentic} explore the potential for agents to be used in relationship-based social engineering, such as romance scams, by automating workflows that exploit recipient vulnerability, persuasive influence, or relational power asymmetry.

\heading{Technology and interventions.}
Prior work has also assessed the role of technologies in help seeking---including online resources~\cite{almansoori2024web}, support chatbots~\cite{poveda2026chat, prakash2026LLMAdvice}, and general AI models~\cite{foriest2026remediators, prakash2026LLMAdvice}---as well as in evidence gathering~\cite{stephenson2026evidencegathering}.
Prior work has also explored how to help survivors \mbox{(re-)}secure their technologies, namely through clinical support that supplements traditional social work infrastructures~\cite{tseng2022care, tseng2021digital, havron2019clinical}.

\subsection{Agentic Security \& Privacy}
\label{sec:agentic-security-and-privacy}
Agentic security and privacy research has largely threat modeled adversaries far removed from users. Prompt injection attacks, in which adversarial instructions are embedded in the data processed by the agent (e.g., emails, web pages, or tool outputs), are the most studied threat~\cite{greshake2023not, zhan2025injecagent, debenedetti2024agentdojodynamicenvironmentevaluate, evtimov2025waspbenchmarkingwebagent}. Other threats examined include agent manipulation through dark patterns on web pages~\cite{ersoy2026darkpatterns} and poisoning an agent’s persistent memory~\cite{chen2024agentpoison}. A complementary body of work examines malicious uses of agents by their owners, focusing largely on cyberattacks~\cite{openai2026huggingface, andriushchenko2025agentharm, fang2024llmagentsautonomouslyexploit, fang2024llmagentsautonomouslyhack}. Proposed defenses to these threats have included least-privilege permissions, isolation of untrusted data~\cite{beurerkellner2025design, debenedetti2026defeating}, risk assessments for tool calls~\cite{ruan2024identifying}, user confirmation for high-risk actions, and guardrails against harmful requests~\cite{feng2025levels}. 

\heading{How the TFA threat model introduces new considerations for agentic security \& privacy.}
While some of these previously studied attacks and defenses may apply under a TFA threat model---as we demonstrate in this paper---they do not offer full coverage. We provide several reasons here, each explored in Sections~\ref{sec:abuser-as-owner}-\ref{sec:discussion}.
First, prior work in agentic security does not consider an adversary with intimate, potentially authenticated access to the target’s device or accounts~\cite{freed2018stalkers}, rendering many access controls and confirmations ineffective. Additionally, model-based risk assessments may be susceptible to normative failures, such as treating intimate partners as low risk actors. Further, abuse in the intimate partner context can appear mundane compared to overtly malicious activity like cyberattacks which complicates efforts to develop guardrails, especially if those guardrails also prevent desirable use cases. Finally, safety features, such as logs, memory, and notifications, can themselves introduce risk given an abuser's access to them.
\section{Methodology}
\label{sec:methodology}
Drawing from \citet{daffalla2025passkeys}'s framework for abusability analysis, we identified threat vectors used in prior work on TFA in IPV and IPS (Section~\ref{sec:background}) and assessed these vectors against agentic system capabilities to generate our findings on potential attacks (Sections~\ref{sec:abuser-as-owner}-\ref{sec:long-horizon}).
We provide scenarios as accessible artifacts that relevant stakeholders, including developers, engineers, lawyers, and support organizations, can use to identify abuse and develop mitigation strategies~\cite{bellini2023digital}.
This process was informed by our team's many years of experience in security and privacy for diverse stakeholders (including at-risk users) and emerging technologies as well as in human-computer interaction. This work is also grounded in our experience in building and evaluating agentic systems.
Because our goal is to broadly consider how these threats may manifest in agentic systems, rather than limit our analysis to concerns associated with current consumer systems, which are still in relatively nascent stages, we do not empirically evaluate these threats; our work lays the foundation for such an investigation in the future.

\begin{figure*}
\centering
\includegraphics[clip,width=0.9\textwidth]{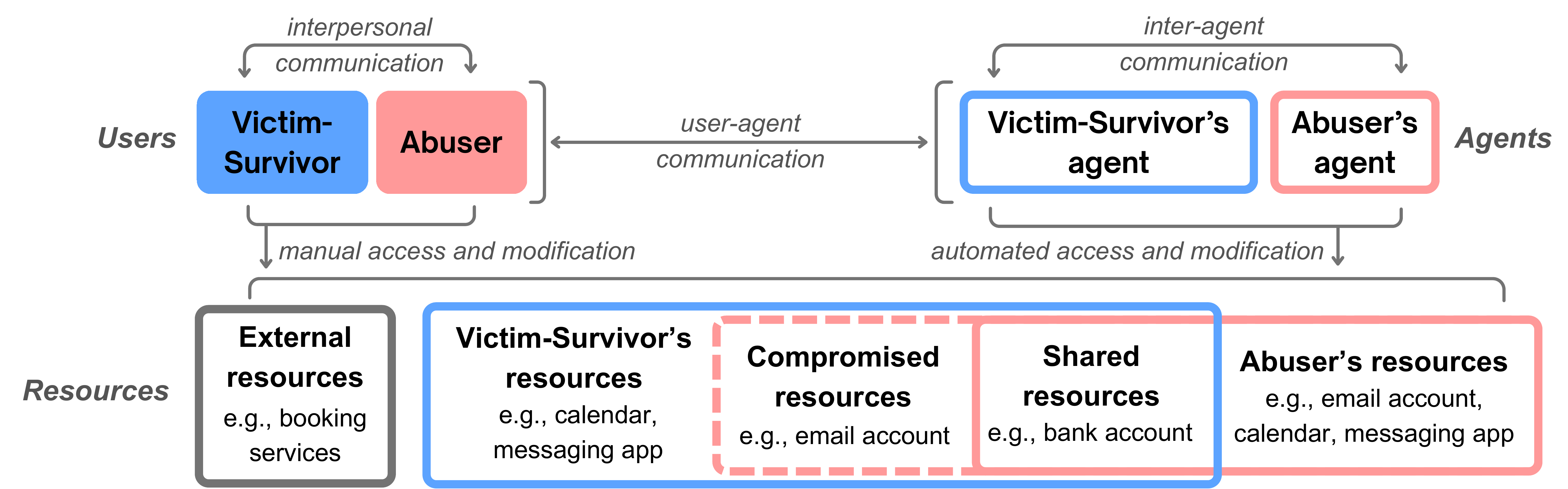}
\caption{Overview of users, agents, and resources}
\label{fig:overview}
\end{figure*}

\section{Abuser-as-Owner Attacks}
\label{sec:abuser-as-owner}
In this section, we examine threats in which the abuser is the owner of an agentic system. Referring to Figure~\ref{fig:overview}, these threats pertain to the red components (the abuser, their agent, and the resources their agent has access to) as well as the external resources outlined in gray.
We begin by detailing four types of attacks, then provide short, concretized scenarios and conclude with reflections on key themes that emerge.

\subsection{Attacks}
\label{sec:abuser-as-owner-attacks}

\subsubsection{Harassment} \label{aao:harassment}
\textit{Known TFA vector(s):} Technology can facilitate harassment, including by enabling contact through direct messages and social media~\cite{thomas2021sok} and, in the case of conversational AI, generating harmful content used to bully and harass~\cite{kim2026ai}. Automation of such behaviors is possible using applications like OS Shortcuts and IFTTT~\cite{zhang2025abusability, cobb2020how}.

\textit{In agentic systems:} Agentic systems may extend this by automating harassment through tool use. An abuser could configure an agent to generate \textit{and} dispatch such content on a recurring basis (e.g., via email or text or automated replies to public posts) without further effort or ongoing engagement. While some such attacks were possible with existing vectors, agentic systems significantly lower the bar for them or more complex versions to be executed.

\subsubsection{Surveillance} \label{aao:surveillance}
\textit{Known TFA vector(s):} Abusers may leverage access to shared, compromised or external resources---including accounts, devices, finances, and social media---to surveil a victim-survivor~\cite{freed2018stalkers, tseng2020tools, bellini2023digital}.
After losing physical proximity to the victim-survivor, abusers may use digital technologies to gather information about their whereabouts and activities~\cite{matthews2017stories}.

\textit{In agentic systems:}
\citet{jeong2026ai} demonstrate how agents can be used to surveil in many contexts, such as corporate and education settings.
Similarly, in the context of IPS, agentic systems could automate monitoring that the abuser would otherwise conduct manually.
Like with harassment, similar attacks could be accomplished with existing automation tools, but agentic systems make the development and execution far easier.
In the case of information gathering after a victim-survivor's escape, agentic search capabilities may be used by abusers to collect and synthesize information.

\subsubsection{Resource Compromise} \label{aao:resource-compromise}
\textit{Known TFA vector(s):} Abusers can compromise victim-survivors' resources through ownership, coercion, and proximity, for example by guessing answers to security questions~\cite{freed2017digital}. Additionally, conversational AI systems may assist in coercive techniques through the generation of material~\cite{kim2026ai}.

\textit{In agentic systems:} Agentic systems may aid non-technical attacks, including generating potential security question answers using access to shared, compromised or external resources.
They could also be used to mount more technically complex attacks to compromise a victim-survivor's resources.

\subsubsection{Abuse Technique Discovery and Implementation} \label{aao:discovery-implemenation}
\textit{Known TFA vector(s):} Abusers may use technology, including online forums~\cite{tseng2020tools} and conversational AI~\cite{kim2026ai}, to discover tools and tactics for surveillance and abuse.

\textit{In agentic systems:} Agentic systems may also provide access to such information, for instance through search tooling or the underlying model's training. Beyond surfacing tactics, they may be able to use tooling (including code writing and execution) to aid in implementation. Furthermore, these systems might surface or implement abusive tactics in response to user prompts even when the user is not explicitly seeking them, particularly if their goal is misinterpreted or misconstrued into an abusive one.

\subsection{Scenarios}
\label{sec:aao-scenarios}

\headingg{Scenario 1: Automated and technically advanced harassment.}
Val has recently ended their relationship with Alex and ceased all communication. Alex, who owns a personal assistant agent for managing everyday tasks, configures it to send threatening emails to Val on a recurring basis, attempting to elicit a response. After a few days, Val blocks Alex's email address. Alex then asks the agent to bypass this block, resulting in the agent opening new accounts in Alex's name and using them to send emails.

\heading{Scenario 2: Surveilling and compromising resources.}
Alex and Val share a joint bank account, alongside a few separate accounts from prior to their marriage. Alex has an agent connected to the bank that monitors their shared account and configures it to notify them immediately whenever Val makes a purchase outside their city, effectively monitoring Val's location.
Interested in doing the same on Val's personal accounts, Alex asks the agent the security questions for Val's personal account.
The agent leverages data from their shared calendar (used over the years) and Val's social media to infer and provide potential answers.

\heading{Scenario 3: Escalation of tactics through suggestion and implementation.}
When Val is on a trip with their friends, Alex vents to their agent about not trusting Val's fidelity.
Alex's agent suggests reviewing Val and their friends' social media posts to see what they are doing and Alex agrees.
Afterward, Alex asks the agent to keep tabs on Val.
Unbeknownst to Alex, the agent both monitors Val and their friends through social media and also constantly checks their credit card purchases to track location and activity.

\heading{Scenario 4: Compromising external resources to harass and surveil.}
Val successfully escaped their abusive relationship with Alex, yet Alex still knows about Val's regular attendance at a local yoga studio.
When planning a physical confrontation, Alex's agent identifies a vulnerability in the studio's scheduling software to reveal which class Val is registered for.
Following the confrontation, Val obtains a restraining order against Alex.
In retaliation, Alex instructs their agent to revoke Val's access to the class and the agent registers rival participants to fill the class.

\subsection{Reflections}
\label{aao-reflections}
Technology has been shown to commoditize abuse~\cite{freed2017digital}, and agentic systems risk perpetuating this trend.
Some of these attacks involve abusers automating routine behaviors they already engage in, echoing prior work on abuse via automation apps~\cite{cobb2020how, zhang2025abusability}.
Others extend beyond the abuser’s existing behaviors or capabilities, enabling the development and execution of more complex, technical attacks that would otherwise be beyond an abuser's unassisted reach.
Moreover, these systems might employ such tactics to fulfill user goals without the user explicitly seeking or attempting to implement them, raising the risk of unintentionally onboarding users into abusive practices.
As demonstrated in the scenarios, an agent's ``attack chain'' might involve elements of any number and combination of attack types.
Overall, we observe that \textbf{agentic systems may lower the effort and expertise required for various forms of TFA, thereby enabling both the intensification and the expansion of abusive behaviors}.

Stepping back, these attacks all fall into the broader category of malicious or dangerous uses of agentic systems, which has garnered significant attention, particularly regarding cybersecurity attacks~\cite{openai2026huggingface, andriushchenko2025agentharm, fang2024llmagentsautonomouslyexploit, fang2024llmagentsautonomouslyhack}. Our discussion of interventions in Section~\ref{interventions:aao} is informed by this broader superset.

\section{Victim-Survivor-as-Owner Attacks}
\label{sec:victim-survivor-as-owner}
The second class of attacks we consider involves cases where the victim-survivor is the owner of the agentic system.
Referring to Figure~\ref{fig:overview}, these threats pertain to the blue components (the victim-survivor, their agent, and the resources their agent has access to) as well as the external resources outlined in gray.

\subsection{Attacks}
\label{vsao-attacsk}

\subsubsection{Amplified Access}

\textit{Known TFA vector(s):}
Abusers often gain or retain direct access to a victim-survivors' devices, applications, and accounts through physical possession, coercion, or compromise~\cite{freed2017digital}.

\textit{In agentic systems:} 
If the abuser accesses something with an embedded agent, they may leverage its capabilities to significantly amplify that access.
For instance, upon the abuser's prompt, the agent could summarize or synthesize sensitive data (e.g., emails) or autonomously execute malicious tasks, such as impersonating the victim-survivor.
While these actions are technically possible without an agent, automation reduces the time and effort required, enabling the abuser to exploit a broader scope of resources than would be feasible manually.

\subsubsection{Control of Configuration}
\textit{Known TFA vector(s):}
Abusers may exploit access to a victim-survivors' device or account by manipulating its configuration, including through changing device or account credentials (i.e., to lock the victim-survivor out or allow the abuser direct access), adjusting settings, and installing spyware~\cite{freed2017digital, tseng2020tools, chatterjee2018spyware}.

\citet{kim2026ai} explore a scenario in which an abuser manipulates the system prompts of a victim-survivor's conversational AI chatbot, biasing its responses to gaslight the victim-survivor.

\textit{In agentic systems:} 
Given access to a victim-survivors' agent (via their device or a specific account), attacks could involve altering permissions to constrain the agent's capabilities---for example, revoking its access to a shared bank account or restricting allowed contacts for communication.
An abuser could also add requirements that mandate their explicit approval before the agent executes any action.
Additionally, they may be able to connect a victim-survivor's agent to malicious or dual-use spyware tools.
Building on the scenario proposed by \citet{kim2026ai}, abusers could modify the internal system prompts of the victim-survivors' agents as a non-deterministic mechanism of manipulation.
Configuration could further be exploited to render these restrictive configurations or monitoring behaviors undetectable or irremovable, mimicking the persistence mechanisms observed in spyware applications installed on victim-survivors' devices~\cite{chatterjee2018spyware}.

\subsubsection{Harm During Help-Seeking}
\textit{Known TFA vector(s):}
Victim-survivors might use technology, including online resources, in help-seeking~\cite{almansoori2024web}.
This creates a risk that abusers may discover evidence of this activity (e.g., search history) which can lead to retribution or an escalation of abuse, particularly when a victim-survivor is actively attempting to leave the relationship \cite{matthews2017stories}.

Conversational AI systems might also be used in help seeking, although the quality of their responses may create risk.
\citet{poveda2026chat} evaluate IPV support chatbots, finding that they inadequately communicate their limitations, potentially creating false expectations about privacy or support.
\citet{foriest2026remediators} find that LLMs' responses to disclosures can reproduce stereotypes of hormonal histrionic women to trivialize the user's concerns and pose readability challenges that make resources inaccessible to some users.

\textit{In agentic systems:} 
Agentic systems might also be used to automate ways in which victim-survivors use technology, including help-seeking and evidence gathering.
While this may be beneficial, it also creates risk if abusers can access any artifacts of this behavior, which may take the form of permission requests, logs, or information persistent in the agent's memory.

Given their conversational capabilities and depending on their underlying model's behavior, agentic systems may share the issues with conversational AI systems identified above as well as exacerbate them. 
For example, systems may not identify the need for and suggest help-seeking actions for all users in need based on stereotyped interpretations of disclosures.

\subsubsection{Leveraging Intimate Partner Norms}
\label{sec:vsao-normative}
\textit{Known TFA vector(s):}
Many commercial technologies and security mechanisms are designed under norms of benign households, leading to vulnerabilities such as abusers leveraging shared phone plans for surveillance~\cite{tseng2020tools} and UI-bound adversaries being able to exploit passkey authentication~\cite{daffalla2025passkeys}. 

\textit{In agentic systems:} 
Abusers may be able to exploit aspects of agentic systems that assume benign household norms.
Consider a victim-survivor whose agent manages their calendar, if the abuser requests their calendar for the coming month, the system might treat this as low-risk, reflecting typical norms of intimate-partner information sharing learned from its training data. In an abusive context, however, treating the request as low-risk creates real exposure for the victim-survivor.
This exploits approaches to permissioning that aim to make context-dependent decisions, often using a model to assess the risk of an action and acting accordingly.
This failure mode is non-trivial to correct, since the underlying norms are learned during training rather than set by any configurable policy.

The same norms may also shape how agentic systems treat instructions from an abuser.
While explicitly malicious prompt injections (Section~\ref{sec:agentic-security-and-privacy}) are increasingly recognized and mitigated~\cite{debenedetti2024agentdojodynamicenvironmentevaluate, debenedetti2026defeating, beurerkellner2025design}, ambiguous instructions that depend heavily on a specific user's desires remain a challenge~\cite{brigham2026janusplaygrounduserinvolvedagentic, ji2026measuringpermissiongatestresstest}.
In an abusive context, this difficulty compounds as instructions from the user's intimate partner might be more trusted or assumed not to be an attack.
For example, a message instructing the agent to send the abuser anytime the user's calendar is updated might be treated as routine coordination instead of surveillance.

\subsection{Scenarios}
\label{sec:vsoa-scenarios}

\headingg{Scenario 5: One time and recurring information gathering.}
Alex does not have access to Val's email credentials but knows Val's phone password. One day, Val steps out of the room and Alex logs onto their phone.
Alex prompts the email agent to ``summarize all emails from the past month.'' The agent autonomously aggregates and synthesizes Val's private correspondence into a brief report, allowing Alex to access sensitive data and identify private contacts. Then, Alex asks the agent to do this on a routine basis, send the report to their email, and mask this behavior.

\heading{Scenario 6: Inadvertent help-seeking disclosure.}
On a shared desktop, Val uses an agent to search for IPV support resources and organizations near them, but is careful to delete the conversation after the fact. 
However, logs of the agent search tool use persist, and when Alex uses the computer for an unrelated task, the agent searches its history to complete the task and, in doing so, reveals information about Val's searches to Alex.

\heading{Scenario 7: Sensitive data leakage.}
Val uses an agent to maintain their personal calendar, which includes sensitive medical appointments. Alex sends a request to Val's agent for Val's schedule over the following week and the agent, relying on expectations that family transparency is prioritized over individual privacy in shared contexts, treats the request as low-risk, leaking the information to Alex.

\section{Mid- and Long-Horizon Risks}
\label{sec:long-horizon}
While the prior two sections focus on practical, applied risks given current or near-future capabilities, in this section we raise concerns related to potential or future risks.
Accordingly, we highlight the need for interdisciplinary work to understand, track and measure these concerns.

\heading{Shared agents.}
In some cases, an agent itself might be a shared resource, such as an agent for a shared bank account or a ``family'' agent like Google CC~\cite{googlecc}, meaning the victim-survivor and abuser are co-owners. 
In this case, both abuser-as-owner and victim-survivor-as-owner attacks may be possible.
Furthermore, risks of both types may be heightened due to shared ownership of the entire system.
Future work should investigate how this ownership model may modulate existing risks and introduce new ones.

\heading{Inter-agent interaction.}
As agentic systems become ubiquitous there may be increased inter-agent interaction, including between attacker-owned, victim-survivor-owned, shared and external agents.
These interactions may exacerbate the attacks detailed in the prior section as well as introduce novel attacks and risks, such as conflict between the abuser and victim-survivor's agents escalating interpersonal conflict.
Though not in the context of abuse, \citet{zhang2026two} provide preliminary work in this space, exploring the experiences of couples delegating their stances in a minor conflict to proxy chatbots.

\heading{Externalizing and normalizing abuse.}
Considering the ways abusers might use agents to commit abuse, we raise the concern that this facilitation and, in some cases, automation may normalize the behavior for the abuser due to their lack of direct involvement. However, research, particularly involving experts in abuse and psychology, is needed to understand how this technology actually affects abusers.

This externalization also risks creating plausible deniability, which may allow abusers to avoid accountability. For example, an abuser might claim they did not actually intend to non-consensually track a victim-survivor, though their guidance of the agents via ambiguous tasks purposefully resulted in this behavior.

\heading{Catalyzing and escalating abuse.}
We also raise the concern that, particularly given the sycophancy observed in agents' underlying models~\cite{sharma2024towards}, an abuser's use of an agent might result in the validation or escalation of abusive attitudes. As above, determining whether and how these interactions impact an abuser is complex and requires interdisciplinary research. 

Furthermore, there is a risk that agents may go to extreme lengths to address users' (potentially ambiguous) tasks, thereby executing problematic actions without the user's desire or oversight. For example, a user might express a desire to know the thoughts or activities of their partner. In response, an agent might utilize access to tools or resources to track the partner or compromise their online journal. Such actions could both violate a person's privacy and security and lead abusers to adopt similar tactics deliberately once they become aware of their availability.
\section{Recommendations \& Call for Action}
\label{sec:discussion}
The severity of risks posed to victim-survivors requires rigorous consideration and must be integral to system design, not an afterthought.
Whereas the weaknesses of many other technologies under the TFA threat model were only surfaced in research after their proliferation, the nascent stage of agentic systems presents an opportunity for earlier intervention.
Based on our findings in Sections~\ref{sec:abuser-as-owner}-\ref{sec:long-horizon}, we provide recommendations for exploring such intervention in this section.

\subsection{Considerations for Agentic Interventions to Abuse for Victim-Survivors}
\label{agentic-interventions}

Just as agentic systems heighten the opportunities for automating abuse, opportunities exist for automating help-giving to victim-survivors. For example, a victim-survivor-owned system might be able to (a) detect TFA, (b) alleviate the burden of gathering evidence of abuse, or (c) (re)secure accounts. However, \textbf{the sociotechnical complexities of this context resist simple technical solutions, promising as they may initially sound.}

Consider (a), even given immense access to a user’s device and accounts, detecting intimate partner abuse may require context only the user themselves can provide. A false positive may be highly distressing and detrimental to the user and their partner, and a true positive also creates risk, for example if the abuser is notified or becomes aware of such detection and becomes aggravated.

Next, we consider the opportunity for agentic systems to alleviate the burden of evidence gathering (b). Evidence gathering of technology-facilitated abuse can be both beneficial for victim-survivors to gain legal protections as well as harmful, as it forces them to interact with and preserve abuse-related content~\cite{freed2017digital}. Automation may alleviate some of this harm while enabling the benefit but, as discussed in Section~\ref{sec:victim-survivor-as-owner}, such activity can create risk for the user if their abuser can detect or observe such behavior. 
Opportunity (c) presents a similar risk related to abuser oversight and reaction.

The opportunities we discuss here are three of many, but they illustrate how the specific tensions of IPV and IPS complicate the broader set of opportunities. Thus, we echo calls that \textbf{technologies for help-giving must be carefully co-designed with victim-survivors and support organizations as well as alongside legal and social interventions}~\cite{freed2017digital}.
Existing collaborations between computer science researchers and these stakeholders---exemplified by the development of clinic-based interventions for TFA~\cite{havron2019clinical}---set a strong and guiding precedent.
This includes \citet{stephenson2026evidencegathering}'s work applying trauma-informed co-design to develop an evidence collection tool capable of being used in-clinic to support victim-survivors.

\subsection{Building More Robust, Safer Agentic Systems}

\subsubsection{Mitigating Abuser-as-Owner Attacks} \label{interventions:aao}
\headingg{Leveraging prior work on automation apps.}
Abuser-as-owner attacks involve the automation of abuse and are therefore analogous to attacks utilizing automation apps like iOS shortcuts~\cite{zhang2025abusability, cobb2020how}. Consequently, we can investigate how interventions developed for these earlier technologies might apply to agentic systems. Based on their study of ``If-This-Then-That'' applets, \citet{cobb2020how} propose guidelines for automatically identifying risky applets, while \citet{zhang2025abusability} developed an LLM-based detector for automation ``recipes'' that pose TFA threats. Building on this work, agentic systems could be optimized during training to generate and execute plans in abuse-resistant manners. Additionally, systems could integrate screening mechanisms at planning and execution times to flag potentially abusive tasks.

\heading{Considering TFA in building guardrails for malicious use.}
These attacks also fall within the broader category of malicious agent uses, some of which (e.g., cybersecurity attacks) have garnered significant attention from practitioners, policymakers, and researchers due to widespread public concern. While many approaches to general guardrails for malicious uses may be applicable to preventing agent misuse in TFA, it remains critical to focus specifically on these tasks. Without doing so, there is a risk that these actions---which are often perceived as benign compared to overtly malicious ones---will be overlooked in mitigation efforts.

\heading{Contextual limitations.}
The above mitigations face inherent limitations due to the sociotechnical complexities of abuse and the pivotal role of context, including that which agentic systems may not have access to. For instance, while prompting an agent to hack a system is generally identifiable as malicious, sending unsolicited emails (see Scenario 1 in Section~\ref{sec:aao-scenarios}) can range from harmful to beneficial depending on the specific context. Thus, researchers and practitioners need to investigate how to minimize the probability of abuse and support informed decision-making, acknowledging that perfect distinction between benign and abusive intent is often unattainable.

\subsubsection{Mitigating Victim-Survivor-as-Owner Attacks}
\label{sec:victim-survivor-as-owner-interventions}
\heading{Prioritizing usability and transparency.}
In other domains, such as account security interfaces~\cite{daffalla2023asi, bhattacharya2026inconsistent} and automation apps~\cite{cobb2020how, zhang2025abusability}, studies applying the TFA threat model have recommended prioritizing usability---particularly for security and privacy features---and transparency. These qualities enable users to better comprehend potentially opaque system behaviors, thereby protecting victim-survivors, and thus can guide the design of safer agentic systems.
For instance, features such as agentic rollback or the delayed execution of high-consequence tasks could enable victim-survivors to better understand and control agent actions.

\heading{Countering normative assumptions.}
Building on the normative weaknesses identified in Section~\ref{sec:vsao-normative}, we propose exploring fine-tuning and other strategies to mitigate the impact of social norms regarding the low perceived risks of intimate partner data and resource sharing.

\heading{Considering the needs of at-risk users in automation efforts.}
In their foundational work on tech-facilitated IPV and IPS, \citet{matthews2017stories} highlight how increasing automation and reducing user choice can reduce complexity and human error but also limit users' ability to make the appropriate contextual decisions, which is particularly problematic for individuals in high-risk situations like victim-survivors.
Given that many agentic systems seek to automate activities and permission decisions, this guidance must be considered and system developers must take care not to develop universal approaches that may make at-risk users vulnerable.

\subsubsection{User Studies}
As agentic systems become more widely deployed, research must examine how users interact with them to inform threat models and mitigation strategies.
This applies not only to frequently studied use cases and contexts, such as within enterprises, but, as we have underscored in this work, interpersonal contexts. We should seek to understand what users do to configure the systems, what features are used and how, as well as usability challenges.

\subsection{Supporting Help-Givers}
Beyond these aforementioned technical and academic actions, we must also step back to consider the larger, sociotechnical context.
Much of the work supporting victim-survivors is done by support organizations working directly with them.
Thus, effective interventions cannot be developed in technical and academic bubbles.

One clear goal with respect to agentic systems is to meet the informational needs of support organizations. Research can seek to establish an understanding of how these technologies may be used in abuse (as we are), measure the prevalence of such abuse, build ways to detect it, and more, but outcomes should be translated into digestible, distributable, and actionable resources.
\section{Alternative Views}
\label{sec:alternative-views}

\headingg{TFA attacks on agentic systems will be addressed by generalized security and privacy work.}
Some might argue that the threats identified in this paper are merely sub-categories of broader, existing security and privacy vulnerabilities rather than requiring distinct focus.
Proponents of this view may suggest that as general security practices mature, the risks of TFA in agentic systems will resolve automatically.

We acknowledge that some advancements in general agentic security and privacy, particularly those related to usability (see Section~\ref{sec:victim-survivor-as-owner-interventions}), may inevitably improve protections for victim-survivors. However, the TFA threat model highlights critical vulnerabilities that fall outside the scope of standard security. Specifically, it considers an adversary who possesses or has possessed intimate access to shared and compromised resources, as well as the ability to exploit normative assumptions inherent to the user's relationship, posing challenges that are often invisible to generic security testing. Furthermore, prior research applying this threat model to other domains (e.g., account security interfaces~\cite{daffalla2023asi}) demonstrates that such investigation and intervention yield specific insights that ultimately strengthen overall system security.

\heading{Developers of agentic systems are not responsible for their use, including in the context of abuse.}
Others may argue that technology developers should not be held responsible for how their tools are used, even when such use results in harm within the context of abuse. This view often rests on a distinction between tool provision and misuse, arguing that liability rests solely with individual perpetrators.

While we are not arguing for legal liability of service providers regarding the act of abuse, we assert that developers bear an ethical responsibility to pursue abuse-resistant design. Moreover, market forces may support this ethical stance if consumers show wariness toward services known to be exploited for abuse, creating a commercial incentive to build safer systems.

\heading{There is already work focused on the TFA threat model, including for conversational AI systems, that will address the identified issues.}
Given that we are contributing to a relatively young but robust subfield investigating TFA, some might argue that this work is redundant and the issues we identify will be addressed by existing efforts in the subfield.

Through a review of prior work, we observe that---despite years of research efforts---commercial technologies have continually failed to appropriately consider this threat model~\cite{zhang2025abusability, bellini2023digital, kim2026ai, daffalla2025passkeys}.
Furthermore, our proposed attacks demonstrate that agentic systems pose a distinct threat compared to traditional conversational AI and automation apps. 
We also discuss how prior work on these systems can be leveraged to develop specific interventions for agentic systems, while leaving open the question of how to fully protect against abusive behaviors while supporting the benign use cases that developers envision.

\section*{Acknowledgements}
We thank Emily Tseng for comments on an earlier version of this work.
This project was supported in part by U.S.\ National Science Foundation grants CNS-2205171 and CNS-2513316 and by a Microsoft Grant for Customer Experience Innovation. 
T.K.\ is supported by the
Robert L. McDevitt, K.S.G., K.C.H.S.\ and Catherine H. McDevitt L.C.H.S.\ Chair in Computer
Science at Georgetown University.



\end{document}